## ABSTRACT

The paper aims to mapping the potential vulnerable areas to illegal dumping of household waste from rural areas in the extra- Carpathian region of Neamţ County. These areas are ordinary in the proximity of built-up areas and buffers areas of 1km were delimited for every locality. Based on various map layers in vector formats ( land use, rivers, buil-up areas, roads etc) an assessment method is performed to highlight the potential areas vulnerable to illegal dumping inside these buffer areas at local scale. The results are corelated to field observations and current situation of waste management systems. The maps outline local disparities due to various geographical conditions of county. This approach is a necesary tool in EIA studies particularly for rural waste management systems at local and regional scale which are less studied in current literature than urban areas.

## INTRODUCTION

Illegal dumping is a major environmental problem in Romania due to the poor waste management facilities from rural areas [1]. New EU regulations imposed one the one hand , the closure of rural dumpsites in 2009 and to the other hand, the extension of waste collection services across rural communities. Statistical data concerning the rural dumpsites reflects certain spatial patterns according to geographical conditions on regional an local scale [2],[3]. Spatial implications of non-compliant urban landfills within the same county are outlined in a geographical context using GIS tehniques.[4] Quantitative assessment method of illegal dumping are developed at commune level [5] for a proper analysis of this environmental issue in rural areas. Furthermore, vulnerability of moutains rivers is assessed using as assessment tools quantitative methods and GIS techniques [6]. These approaches argue the role of geographical conditions in illegal dumping field in the context of low coverage of population to waste collection services. The paper proposes an assessment model in order to determine the susceptibility of rural areas to illegal dumping based on GIS techniques.

**METHODS**

GIS techniques are already optimal tools in assessment studies concerning the location of landfills from urban areas in different geographical areas such as [7],[8],[9]. This paper aims to map potential areas vulnerable to uncontrolled disposal of waste from rural areas of extra-Carpathian region of Neamț County based on GIS techniques. These vulnerable areas are commonly found in the proximity of built-up areas inside the buffer area of 1 km. Administrative areas of following cities such as Roman, Târgu Neamț (including villages Humuleşti, Blebea) and Roznov (Chintinici, Slobozia) were excluded from this analysis. Spatial analysis of rural dumpsites and field observations from the period 2009-2011 highlight the fact that the main factors (in the proximity of villages) susceptible to waste dumping are: rivers /creeks; floodplains, pastures, degraded lands ( landslides, gully erosion ), old geological sites ( ex. loam ), local roads, forest areas ( less common). Thus, most of these factors (without forests areas) were extracted inside buffer areas of 1km for every village from study area based on the one hand to general vector layers (rivers, roads, administrative territorial units of communes) and to the other hand, the land use extracted in its turn from Corine Land Cover (those with more than 5 ha surface) or from vectorization process on topographic maps (scale 1: 25 000). Restrictive factors such as altitudine, slope and buffer areas (1 /0,5/0,25 km) must be taking into account for the classification and assessment process of vulnerability to illegal dumping. These factors influence the vulnerability of each susceptible element mapped within the buffer area of 1 km. Also it is used the SRTM (Shuttle Radar Topography Mission) for achieving the maps and their modeling. In order to determine the vulnerability according to altitude and slope firstly these factors were modeled from SRTM and upgrade their levels of altitude and slope as follows:

| Altitude (m) | Class / note |
|---|---|
| <200 | 20 |
| 200-300 | 19 |
| 300-400 | 18 |
| 400-500 | 17 |
| 500-600 | 16 |
| 600-700 | 15 |
| 700-800 | 14 |
| 800-900 | 13 |
| 900-1000 | 12 |
| 1000-1100 | 11 |
| 1100-1200 | 10 |
| 1200-1300 | 9 |
| 1300-1400 | 8 |
| 1400-1500 | 7 |
| 1500-1600 | 6 |
| 1600-1700 | 5 |
| 1700-1800 | 4 |
| 1800-1900 | 3 |
| 1900-2000 | 2 |
| >2000 | 1 |

| Slope | Class / note |
|---|---|
| 1-3 | 8 |
| 3-5 | 7 |
| 5-7 | 6 |
| 7-10 | 5 |
| 10-15 | 4 |
| 15-20 | 3 |
| 20-25 | 2 |
| >25 | 1 |

Fig. 1 Rate of evalution for altitude and slope

Altitude and slope can determine what factors susceptible to waste disposal have an increased vulnerability in the proximity of villages.

Therefore, regions located at low altitudes and low slopes are favorable for the development of human settlements, these areas are also exposed waste dumping. For example, floodplains of rivers from Subcarpathian depressions (which are only used as pasture for livestock) are the most exposed places regarding the uncontrolled waste disposal. On the other hand, in the hilly regions, pastures, roads or degraded lands become options for waste disposal if these sites are not located in higher areas or on higher slopes. Easy and convenient access to these places will increase the vulnerability rate to illegal dumping. The presence of these susceptible factors within buffer of 1 km were noted with the value 1 and those outside buffer with 0. Also, buffers areas was noted according to proximity of built-up areas as follows :

- buffer of 250m - note 3,
- buffer of 500m - note 2
- buffer of 1000 m – note 1

Thus, the proximity (buffers 0.25 / 0.5 / 1km), slope and altitude will determine the most likely places exposed to illegal dumping. The vulnerability gradient is determined on one part, by the sum of total classes /notes ( restrictive & susceptible factors) and to the other part by weighting each factor. Firstly, this weighting performs a hierarchy according to the influence played by restrictive factors, and on the other hand, a hierarchy for susceptible factors according to their frequency on field where such sites became illegal dumps as shown in following formula from figure 2:

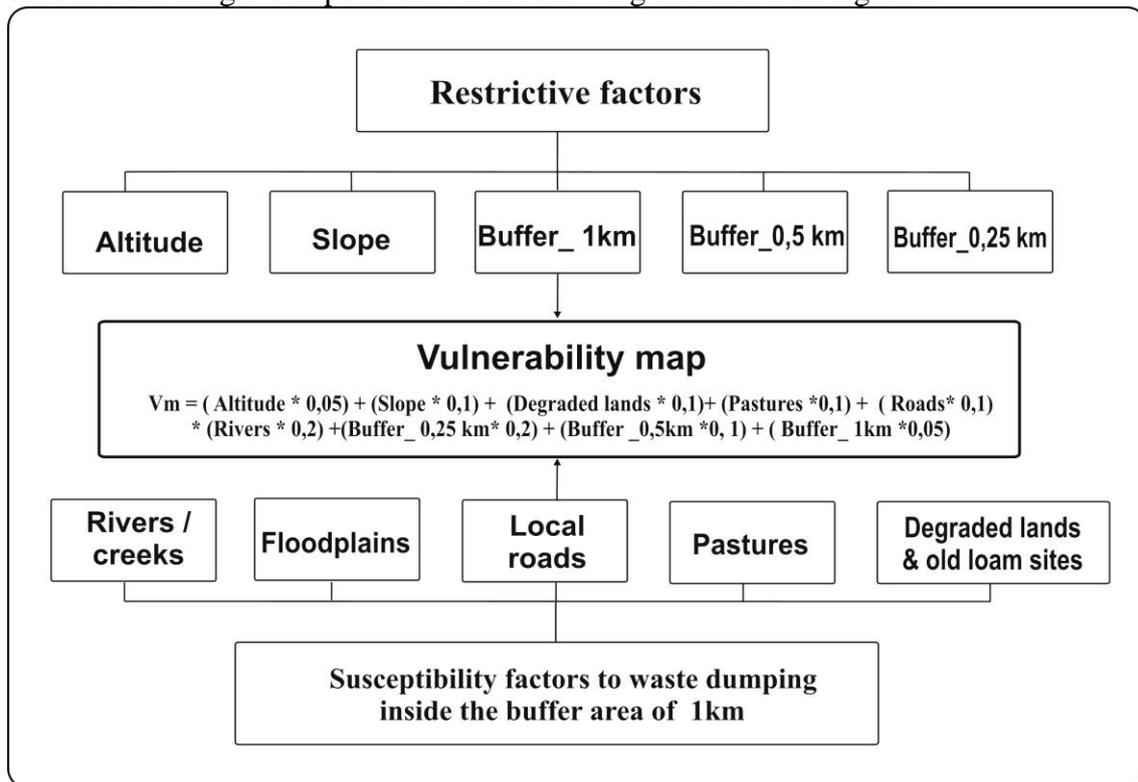

Fig. 2 Methodological scheme and achievement of vunerability map

The resulting values ( 5-39) are divided into 7 classes (ecart 5 points). Each class show a certain category of vulnerabilty to illegal dumping :

- very low ( 5-10) no such areas are determined within buffer areas,
- low (10-15),
- insignificant (15-20),

- moderate (20-25),
- significant (25-30),
- high (30-35)  and
- very high (>35).

The  method is validated by rural dumpsites or illegal dumping points indentified during field observations from 2009 (September) – 2011 and also based on satellite images provided by Google Earth. Therefore, from 163 of such sites located on the map, only 4  were not validated by this method  being located outside buffer of 1km towards built-up areas of villages. Also, no such sites were  found in very low or low areas of vulnerability. Most of these illegal dumps were  found for high class (64), significant (45), very high (25), moderate (22) and more less for insignificant (3). The large number of dumpistes counted for significant and high vulnerability classes is  due to their location on  floodplains of  Siret, Moldova, Bistriţa and Ozana rivers  frequently outside of  buffers  (250  & 500 m). Illegal dumping  is still present although rural dumpsites should be closed and rehabilitated after 16 July 2009. This practice is dispersed across a commune or even village where several such sites are located in a close distance to each other. Field monitoring of uncontrolled waste disposal varies depending on the the geographical context. Thus, the villages in the vicinity  of large rivers (from subcarpathian and coridor valley sector)  dispose  most of  household waste generated into their  floodplains.  On the other hand, this practice is difficult to follow in the case of communes/villages with tentacular structures of built-up areas which are frequently located in the hills regions of county (eg. Bălţăteşti, Grumăzeşti). In this regard, the methodology aims  to emphasize which  factors susceptible to uncontrolled waste disposal are the most vulnerable.

**RESULTS AND DISCUSSION**

Different  geographical  conditions  from  study  area  lead  to  various  options  of uncontrolled disposal of household waste. In contrast to the mountainous region (where geographical barriers mitigate the number of these options, waste disposal is taking place mainly in rivers / creeks or on their banks)  in subcarpathian and plateau sector of county, more local factors are vulnerable to  such bad practices in the proximity of villages. The household waste  is usually disposed  on sites which are not important in terms of the local economy, these sites are  near the household and does not require considerable  effort  to  reach  them  outlining  the  principle  of  ''proximity  and convenience.'' For this reason, the vulnerability map is correlated with land use. On the one hand,  forests areas, arable lands, vineyards, orchards and cultural complex prevail in land use patterns outside the buffer  area of 1km ( blanks areas between buffer areas ) and on the other hand, these land uses are not susceptible to waste dumping in the proximity of built-up areas. However, the forests on the outskirts of villages or along local roads that cross these areas may be vulnerable to such practices [10]. Low or insignificant vulnerability to illegal dumping  generally correspond to the crops fields in the vicinity of these villages. In contrast, rivers and streams sectors within built-up areas or inside  buffer ares of  250 m are the most vulnerable to uncontrolled waste disposal. The major riverbed of a river that passes through the built-up area of a village is the main output of waste generated by household in the proximity.

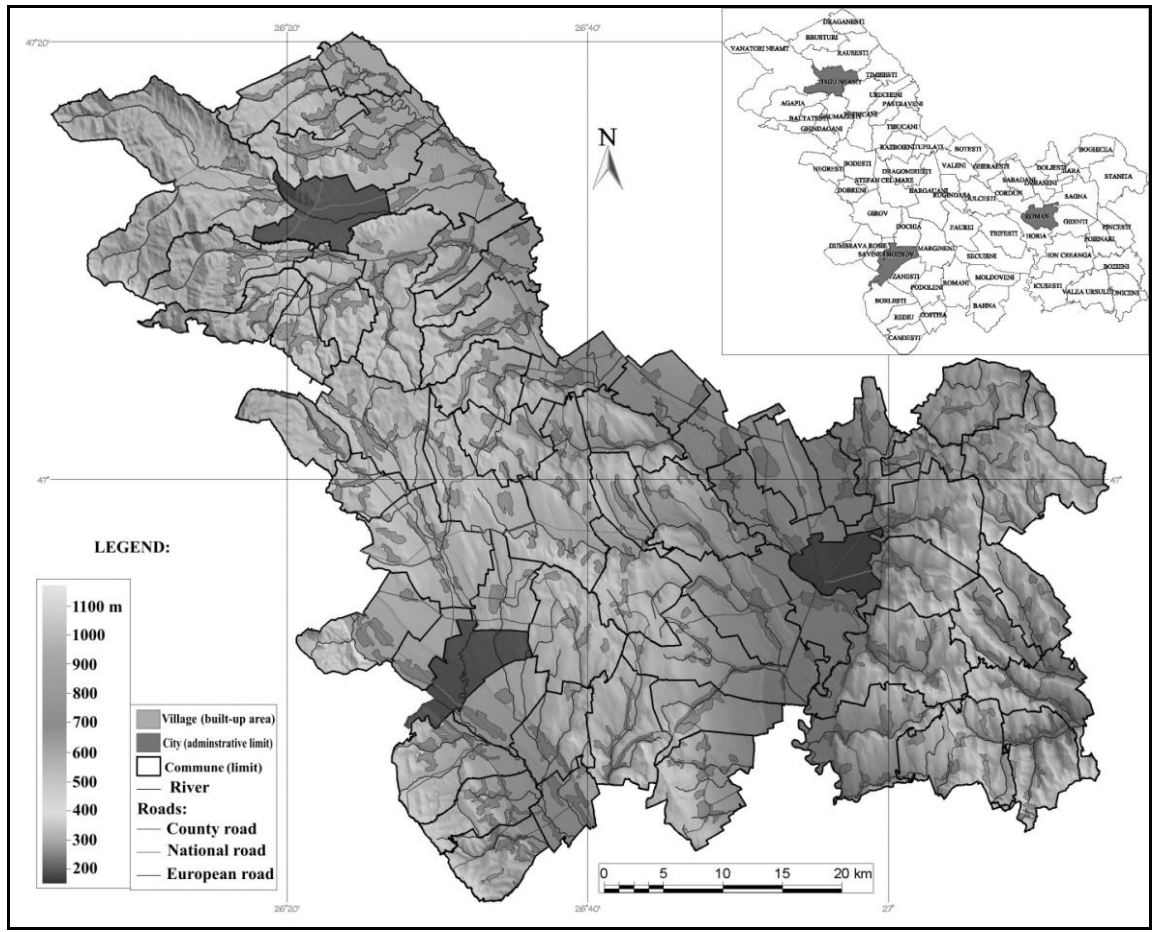

Fig. 3 Extra- Carpathian region of Neamț County

The most common locations are at the intersection of local roads from one bank to another. Inside built-up areas, the creeks beds, embankments, slopes of fluvial terraces, local roadsides (which lead to arable lands) are ordinarily highly exposed to illegal dumping. Ponds or puddles near villages or within built-up areas are also exposed to uncontrolled waste disposal. These water bodies are cleaned in case of floods or swift rises of tributaries encouraging such bad practices outlined also by field observations (Tămășeni, Lutca villages). Pastures or uncultivated lands near the households and / or local roads are most susceptible for open dumps. Local roads and pastures inside buffer areas of 500 m have a significant or high vulnerability to illegal dumping particularly in hilly region of Moldavian Plateau (Boghicea, Stanița and Bâra communes). Of the other part, geomorphological factor is cumulated with the economic one, so degraded lands in the vicinity of villages (due to landslides or erosion) which have no agricultural use are significant susceptible to illegal dumping. Degraded lands likely to be used as open dumps are those from Moldavian Plateau, especially in the upper basin of Bârlad river ( Valea Ursului, Oniceni and Bozieni commuens). Old loam sites are used as local dumps, being a suitable solution than the rest of options in the absence of an organized waste management system. However, these locations are less common in the study area, such disposal sites were found in Botești, Tămășeni, Bârgăoani and Tupilați villages. Frequently, floodplains of rivers in the proximity of villages or creeks within buil-up areas are most vulnerable to illegal waste disposal as shown in figure 4.

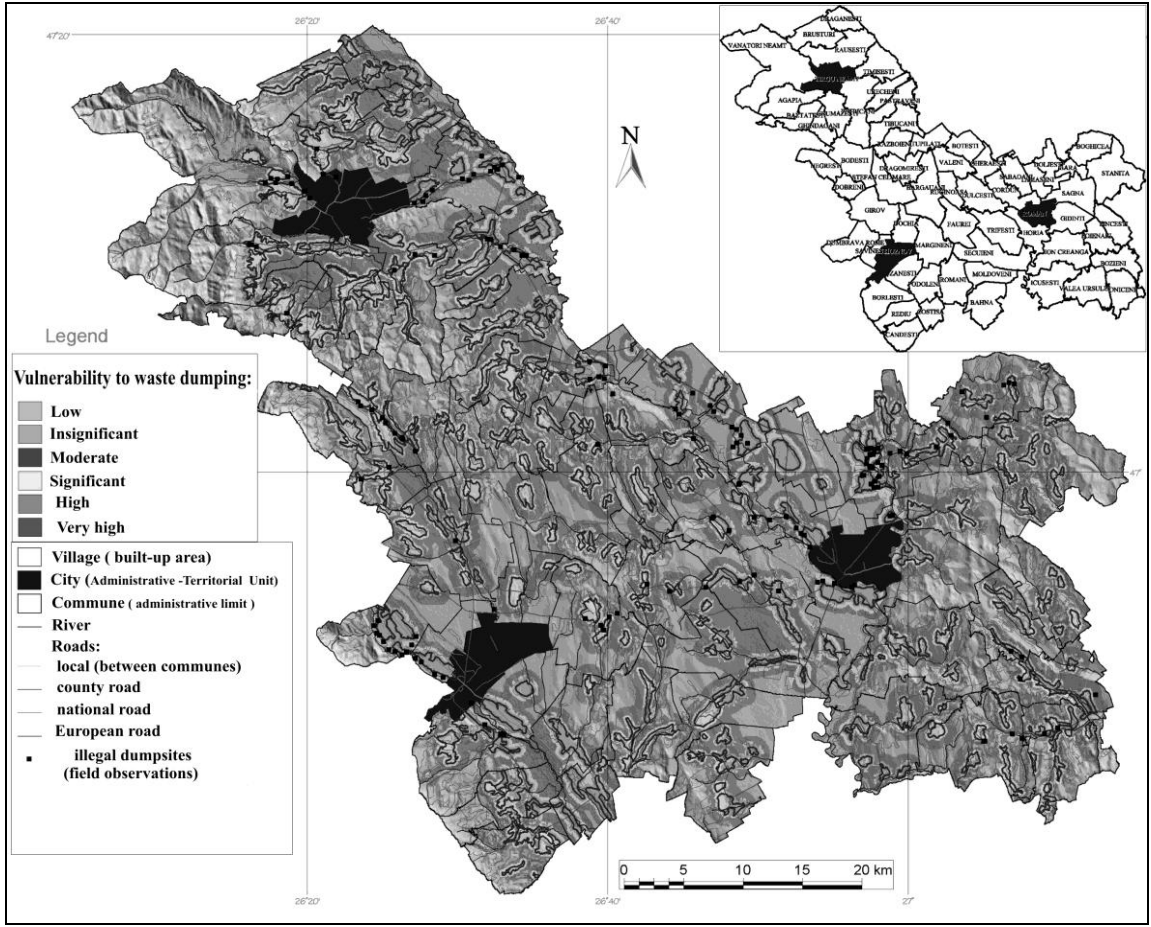

Fig. 4 Vulnerability to illegal dumping in study area

The local population is still unaware of negative implications on local environment and potential destructive effects of these bad practices. The map outline the fact that vulnerability of these areas is generally high or significant for Bistrița, Cracău, Ozana, Moldova and Siret rivers, such dumpsites being also observed on the field or through Google Earh images. Local geomorphological context may play an important role concerning the factors susceptible to illegal dumping. Geographical location of a village, depending on proximity between built-up area and fluvial relief may determine for instance the waste disposal either in a riverbed river / creek , on the floodplain, on slope or on top of river terrace if the distance to the watercourse is growing. These approaches can vary from one commune to another (eg Cordun vs Săbăoani) or from one village to another (Cut vs Izvoare for Dumbrava Roşie commune). Wider floodplains became favorite locations of local communities in uncontrolled waste disposal because the dumpsites located in these areas are regularly taken by stronger floods.

**CONCLUSIONS**

GIS techniques outline the fact that local geographical conditions from study area influence the gradient of vulnerability to waste dumping. The most exposed lands are the susceptible factors in the proximity of built-up areas inside the buffer areas of 250 & 500 m.

These factors vary one the one hand in different geographical areas such as subcarpathian depressions and hills, corridor valleys or Moldavian Plateau and on the other hand, at local scale from one locality to another. This approach highlights the role of restrictive factors such as proximity, slope an altitude in the determining process of the most vulnerable sites to illegal dumping. The modeling method of vulnerability was validated by several such waste disposal sites indentified on the field. This method is a necessary tool for EIA studies applied in rural areas and it is a complementary support for quantitative assessment methods of illegal dumping.

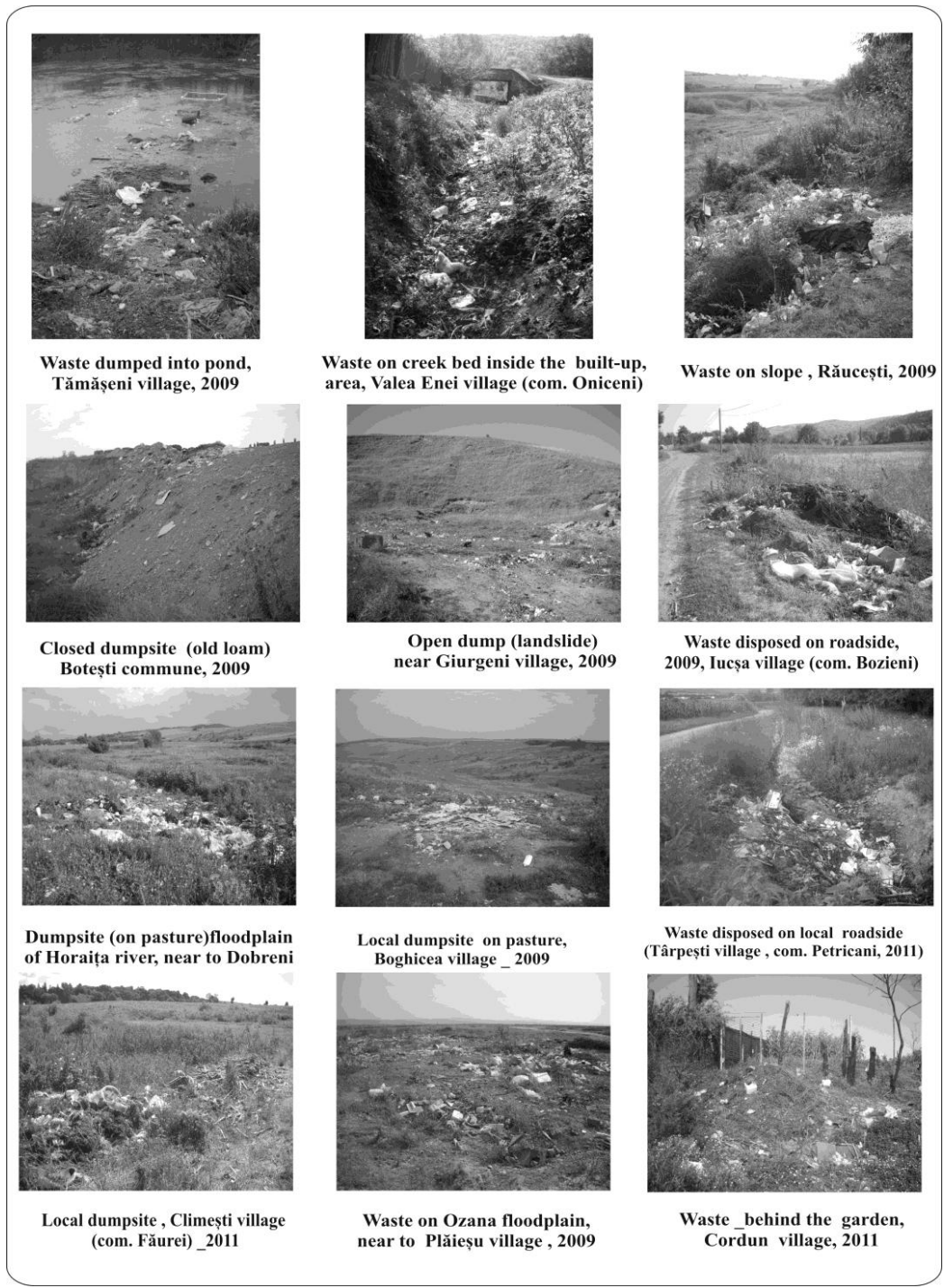

Fig. 5    Illegal dumping of waste in rural areas